\documentclass[lettersize,journal]{IEEEtran}
\usepackage{amsmath,amsfonts}

\usepackage{array}

\usepackage{textcomp}
\usepackage{stfloats}

\usepackage{verbatim}
\usepackage{graphicx}

\usepackage[hyphens]{url}
\usepackage[hidelinks]{hyperref}
\usepackage{xcolor}
\usepackage{cite}
\usepackage[ruled]{algorithm2e}
\usepackage{algorithmicx}
\usepackage{mathptmx}
\usepackage{caption}
\usepackage{subcaption}
\usepackage[capitalize]{cleveref}
\begin{document}

\title{Hierarchical Control for Vehicle Repositioning in Autonomous Mobility on Demand Systems}

\author{Pengbo Zhu, Giancarlo Ferrari-Trecate, Nikolas Geroliminis
\thanks{Pengbo Zhu and Nikolas Geroliminis are with the Urban Transport Systems Laboratory, \'Ecole Polytechnique F\'ed\'erale de Lausanne, 1015 Lausanne, Switzerland (email:  {\tt\small pengbo.zhu@epfl.ch,  nikolas.geroliminis@epfl.ch}).

Giancarlo Ferrari-Trecate is with the Dependable Control and Decision Group, \'Ecole Polytechnique F\'ed\'erale de Lausanne, 1015 Lausanne, Switzerland (email: {\tt\small giancarlo.ferraritrecate@epfl.ch}).}}



\maketitle

\begin{abstract}

Balancing passenger demand and vehicle availability is crucial for ensuring the sustainability and effectiveness of urban transportation systems. To address this challenge, we propose a novel hierarchical strategy for the efficient distribution of empty vehicles in urban areas. The proposed approach employs a data-enabled predictive control algorithm to develop a high-level controller, which guides the inter-regional allocation of idle vehicles. This algorithm utilizes historical data on passenger demand and vehicle supply in each region to construct a non-parametric representation of the system, enabling it to determine the optimal number of vehicles to be repositioned or retained in their current regions without modeling the system. At the low level, a coverage control-based controller is designed to provide inter-regional position guidance, determining the desired road intersection each vehicle should target. With the objective of optimizing area coverage, it aligns the vehicle distribution with the demand across different districts within a single region.
The effectiveness of the proposed method is validated through simulation experiments on the real road network of Shenzhen, China. The integration of the two layers provides better performance compared to applying either layer in isolation, demonstrating its potential to reduce passenger waiting time and answer more requests, thus promoting the development of more efficient and sustainable transportation systems.

\end{abstract}

\begin{IEEEkeywords}
Autonomous Mobility-on-Demand systems, hierarchical control, data-driven control, vehicle rebalancing, taxi fleet control
\end{IEEEkeywords}

\section{Introduction}
\IEEEPARstart{T}{he} continuous expansion of modern city areas and worldwide increasing population density in megacities is leading to the rapid growth of people’s mobility demands. Thus, Mobility-on-Demand (MoD) system (such as Uber, Lyft, and Didi) is a promising solution to provide passengers with efficient and fast service by deploying a group of coordinated vehicles serving people’s requests for rides within city areas. 
Particularly, using autonomous vehicles as MoD (Autonomous Mobility-on-Demand, AMoD) fleets brings notable advantages, such as increasing the available driver supply to meet rising demand and significantly reducing transportation costs \cite{Zardini2022}. 

The imbalance of demand and supply can compromise the system efficiency. One key factor is the asymmetry in the origin and destination distributions of trips, which creates discrepancies between actual and desirable spatial distributions of MoD fleets. For example, when vehicles only cruise freely without cooperation, some districts can be oversupplied, whereas customers who need a ride in other regions cannot be served due to a lack of available vehicles around. Moreover, on-demand services with larger fleet sizes and numerous idle vehicles can adversely impact passenger mobility. This is due to a substantial increase in empty kilometers traveled, which also leads to traffic congestion \cite{Caio2021}. To mitigate these challenges, the proactive relocation of idle vehicles to high-demand areas in real-time—known as vehicle rebalancing/repositioning—is a crucial strategy that can significantly improve MoD system performance \cite{Alex2018,Marczuk2016}. In \cite{Pavone2010}, the urban road network is partitioned into smaller zones, denoted as stations/regions. A rebalancing policy is implemented to manage the transfer of empty vehicles across these stations.
A fluidic discrete-time approximation model is introduced and an optimization approach is presented for steering the system to equilibrium \cite{Pavone2012}. Model Predictive Control (MPC) approaches are employed to determine the optimal rebalancing policy in \cite{Andrea2019,Mohsen2018}, while \cite{Zhang2016} provides proof of its stability in the sense of Lyapunov. Both historical and real-time data are utilized in \cite{Miao2016} to build a prediction model. Furthermore, in \cite{AlonsoMora2017B}, the problem is described as an integer linear programming problem that dispatches vehicles to locations with recent unfulfilled passenger requests.

\subsection{Related works}

 \textbf{Hierarchical framework} In the context of vehicle rebalancing, most works involve initially dividing the city into smaller regions, therefore they focus on determining the number of vehicles that should transfer from one region to another. However, advancing beyond this region-level scope to provide finer-grained, such as position-level guidance, remains a challenging problem. 
In a large-scale urban transport network, it is intractable to model microscopic traffic components due to its high complexity issues. A widely-used approach to solving large-scale system problems is to partition the system into smaller subsystems first and design a multi-layer control scheme. It benefits from efficient coordination between the actions of upper-level controllers which operate the aggregated traffic components, and the self-management of individual vehicles at lower level \cite{Schutter2006}. In \cite{sirmatel_tsitsokas_kouvelas_geroliminis_2021}, a hierarchical control architecture in traffic management problem was studied, where the higher-level regulator for coordination and optimization is interfaced with lower-level controllers acting at the local scale. A hierarchical control of heterogeneous large-scale urban road networks is presented for efficient use of network capacity in \cite{YILDIRIMOGLU2018}.


\textbf{Data-driven methods} While MPC has been proven an efficient tool for traffic management of large-scale transportation systems (see for example perimeter control in multi-region urban networks with Macroscopic Fundamental Diagram (MFD) representation in \cite{ramezani_haddad_geroliminis_2015}), deriving a dynamic model of the system is a prerequisite which heavily relies on expert experience and system knowledge. Moreover, given the complexity and non-linearity inherent in modeling such systems, it can lead to issues in modeling or adapting to different scenarios. For example, a detailed aggregated network model for ride-hailing systems was developed in \cite{Caio2021}. However, integrating this model into numerous regional control systems may be challenging, as the number of variables grows with the number of regions considered. 

Nowadays, data-driven approaches receive more and more attention, especially when the system is too complex (for example, for applications with humans in the loop), or too costly to model thoroughly and identify necessary parameters. Data-driven methods enable the learning of control policies directly from data. A data-driven distributionally robust method is proposed in \cite{Miao2021} to minimize the worst-case expectation cost meanwhile considering the uncertainty demand. A deep reinforcement learning approach for rebalancing was investigated in \cite{JIAO2021} to improve the income efficiency of drivers. For shared MoD systems, a model-free reinforcement learning scheme is proposed to offset the imbalance in \cite{Jian2017}, however, it has the risk of dispatching more vehicles than needed to an area. Moreover, reinforcement learning methods usually need huge amounts of data to yield good performances, and safety constraints are not considered in most of these works. In contrast, inspired by behavioral models of linear systems \cite{MARKOVSKY2005, Markovsky2006}, a Data-enabled Predictive Control (DeePC) algorithm is presented in \cite{Coulson2018}, which not relying on a parametric system representation. Using real-time feedback, DeePC drives the unknown system along a desired trajectory while satisfying system constraints. By assuming measurable disturbances, DeePC has been implemented to mixed traffic flow and power systems\cite{Wang2022, Huang2022}. The robustness of regularized DeePC for stochastic and nonlinear systems was discussed in \cite{Coulson2019, Berberich2021} and \cite{Dorfler2021}. Our earlier work \cite{Pengbo2023ECC} showed that a DeePC-based controller could help the fleet management to balance supply with demand at the upper level via inter-regional vehicle transfers.



\textbf{Coverage Control methods}
In the past several decades, coverage control algorithms have played an important role in robotic systems due to their capability to coordinate and control multiple mobile agents simultaneously \cite{Sonia2011}. In \cite{Sotiris2016,Sotiris2018A,Sotiris2018B}, the authors focus on methods to deploy the mobile agents to maximize the total covered area. Centroidal Voronoi configurations \cite{Qiang1999}, the generators of which coincide with the center of mass (i.e., the centroid) of Voronoi cells, are widely used in location optimization problems. Lloyd’s algorithm \cite{Lloyd1982} provides a tractable approach to compute the centroidal Voronoi configuration and steers agents towards the centroids of their own cells. 
Considering a practical setting, \cite{Yun2012} and \cite{Durham2012} use graph Voronoi partition \cite{Erwig2000} to represent the non-convex environment. Coverage control is applied to control empty vehicles across the whole urban area in MoD systems in \cite{Pengbo2022} and shows its advantage in reducing passenger waiting time. Motivated by the great potential in applying coverage control to demand-responsive fleet rebalancing, we adopt it as a lower-level controller in our framework. It can be operated in a distributed way where each vehicle computes its own desired intra-regional position, i.e., which intersection it should move towards. 


\subsection{Contribution}

Building upon our earlier conference work \cite{Pengbo2023ECC}, 
this paper provides the following contributions:

1) Different from \cite{Pengbo2023ECC}, we introduce a hierarchical structure attempting to balance empty vehicle supply and passenger demand efficiently. The framework consists of multiple control levels: the upper level manages inter-regional vehicle relocation in a centralized manner, while the lower level provides position guidance for individual vehicles. In this paper, we implement a Voronoi-based coverage control algorithm in the lower level to align the vehicle distribution with the demand density within each subregion. This hierarchical structure bridges the gap between macroscopic and microscopic perspectives, demonstrating advantages over applying each layer solely.

2) To the best of our knowledge, this is the first study to apply a data-enabled predictive control algorithm \cite{Coulson2018} to the problem of real-time fleet rebalancing. This approach circumvents the complexities associated with MoD system modeling and identification.

3) This work extends \cite{Pengbo2023ECC} also in the following aspects: (i) We conduct a sensitivity analysis on the design of the hyperparameter in the control objective, offering insights into its impact on system performance. (ii) We assess the robustness of this data-driven method by introducing noise into the data, which is a condition commonly encountered in real-world scenarios. This test further highlights the reliability of this method in practical settings.

The remainder of this paper is organized as follows: in \cref{sec: hierarchicalcontrol}, an overview of the hierarchical control framework for vehicle rebalancing is provided. The DeePC-based upper-level controller is introduced in \cref{sec: upper}. Then we present the coverage control-based lower level controller and summarize the implementation combining both levels in 
\cref{sec: lower}. In \cref{sec: results}, we test the proposed method on an AMoD simulator using the real city road network of Shenzhen, and compare it to other policies. We conclude the whole work and some future work is described accordingly in \cref{sec: conclusion}.
\begin{figure*}[htp]
    \centering
    \includegraphics[width=0.78\textwidth]{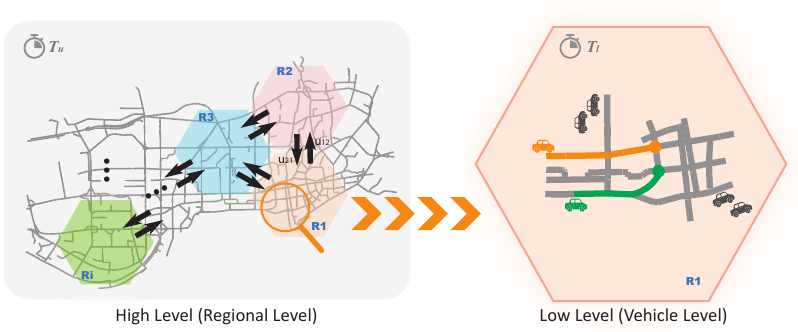}
    \caption{Hierarchical control structure for vehicle rebalancing in AMoD system. The inter-regional vehicle transfers, determined by the high-level controller, are represented by black arrows. On the right, these transfers are depicted as black taxis, which are directed to relocate to their desired regions accordingly. The orange and green trajectories on the right indicate the intra-regional position guidance for idle vehicles, as given by the low-level controller.}
    \label{fig:hierarchical}
    \vspace{-0.4cm}
\end{figure*}
\section{Hierarchical control for Repositioning}\label{sec: hierarchicalcontrol}

A centralized global controller for large-scale systems may suffer from scalability and computational complexity issues. A well-known and effective solution is to partition/cluster the systems into small-scale, non-overlapping subsystems \cite{Lakshmikantham1981}. The higher-level controller addresses more abstract control problems at a slower timescale. It can function as a centralized controller collecting aggregated information across all clusters. The lower-level controller, operating at a faster timescale, can be designed in a distributed manner, leveraging local information or measurements to optimize local dynamics. Notably, controllers at different levels can be tasked with different objectives \cite{Schutter2006}. 

The proposed hierarchical structure for vehicle rebalancing is illustrated in \cref{fig:hierarchical}. In this model, the upper-level aggregates data from all regions, e.g., the number of requests issued and the count of empty vehicles per region. The control action generated at this level specifies how many vehicles should remain in their current regions or be moved to other regions, to balance current and future passenger demands with vehicle supply. The lower level, in contrast, accesses real-time detailed information, such as the coordinates of each vehicle and its occupancy status. For empty vehicles that remain in the current subregion, they communicate and collaborate with one another to achieve a position configuration aligned with the demand density within the region. 

Particularly in this work, our focus is on designing the upper controller through a regularized Data-Enabled Predictive Control algorithm with the dual aim of maximizing the number of requests fulfilled while considering the rebalancing cost; A coverage control algorithm functions as the lower controller, to achieve a good intra-region vehicle position distribution to reduce the average waiting time. Important variables in this AMoD system and their descriptions are provided in \cref{tab:variables}.


\textit{Notations: } 
Unoccupied vehicles can be classified into two types: `relocating vehicles', which are instructed to relocate to another region, and `idle vehicles', which remain in their current region. Collectively, both types are referred to as `empty vehicles'. 

\begin{table}[htb]
\centering
\begin{tabular}{|c|c|}
\hline
\textbf{Symbol} & \textbf{Description} \\ 
\hline
$k$ & discrete time index for the upper layer,\\
&using a sampling interval $T_u$\\
\hline
$I, J$& region index\\
\hline
$u^{IJ}_k$        & input variable: number of vehicles that should\\
&move from Region $I$ to Region $J$\\
\hline
$y^I_k$& output variable: number of answered requests \\
&starting from Region $I$\\
\hline
 $w^{O,I}_k, w^{D,I}_k$&  disturbance variable: number of\\ 
& passenger requests starting from \\
&and ending in Region $I$, respectively \\
\hline
$\Tilde{w}_k$ & ground truth value of $w_k$\\
\hline
$e_k^I$& number of empty vehicles in Region $I$ \\
&at time step $k$\\
\hline
$\theta^{IJ}_k$ & vehicle flow transfer ratio from Region $I$ to Region $J$\\
\hline
$k'$ & discrete time index for the lower layer,\\& using a sampling interval $T_l$\\
\hline
\end{tabular}
\caption{Notation for variables}
\label{tab:variables}
\end{table}
\section{Data-driven upper-layer control}\label{sec: upper}
To address the imbalance between vehicle supply and customer demand across different regions, in this section, we use a data-driven rebalancing strategy that guides empty vehicles to move to another region/stay in the same region.
We first present the problem formulation for the upper-layer control, followed by a detailed implementation of Data-enabled Predictive Control (DeePC) for instructing inter-regional vehicle transfers.



\subsection{Input, output and disturbance in AMoD systems}\label{subsec:motivation_upper}
\begin{figure}[htb]
    \centering
    \includegraphics[width=0.33\textwidth]{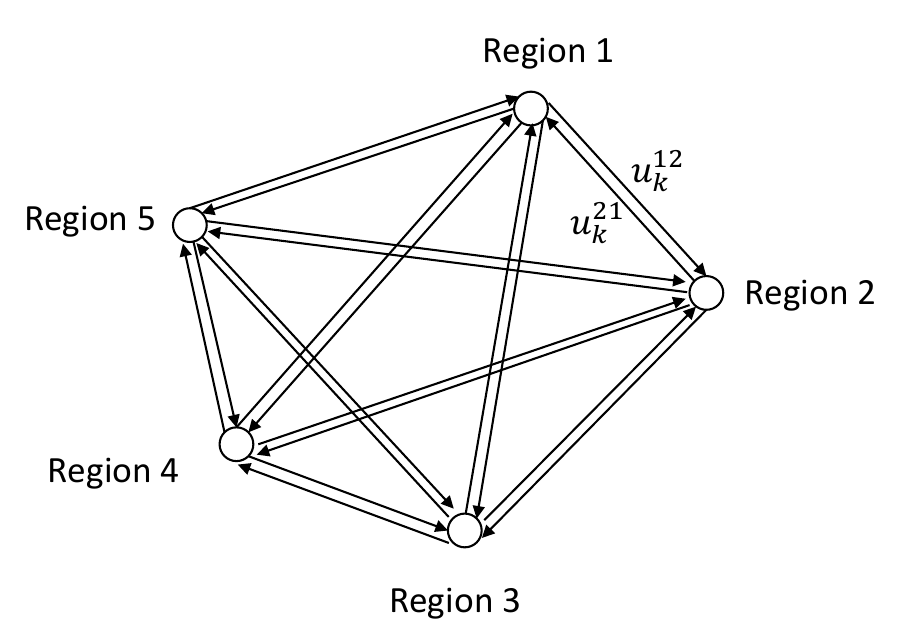}
    \caption{A schematic diagram of vehicle rebalancing. A five-region example case is shown and a centralized controller gives region-level transfer guidance to empty vehicles in each region, e.g., $u_k^{12}$ informs how many vehicles in Region 1 are asked to relocate to Region 2 at time $k$.}
    \label{fig:regionTransfer}
\end{figure}
\cref{fig:regionTransfer} shows a schematic diagram of the rebalancing task in an urban area partitioned into $R$ regions. The control input $u_k$ produced by the upper-layer controller is defined as $u_k = [u^{11}_k, u^{12}_k, \dots, u^{1R}_k, u^{21}_k, u^{22}_k,\dots,u^{2R}_k,\dots, u^{R1}_k, u^{R2}_k, \dots, u^{RR}_k]^T$, and collects variables $u^{IJ}_k$, each specifying how many vehicles should move from Region $I$ to Region $J$ (see also \cref{tab:variables}).
 Particularly, $u^{II}_k$ stands for the number of vehicles that should stay in the current region.

Taxi network companies, such as Uber and Lyft, can access and collect data from both the demand side (position and time information of requests, etc.) and the supply side (vehicle positions and occupied status, etc.). The passenger demand is uncontrollable but measurable, i.e., it is a disturbance variable. We use the subscript $O$ to denote `Origin' and $D$ for `Destination'. For defining stacked vectors $a_1,a_2,\dots,a_i$, we use the notation as $col(a_1,a_2,\dots,a_i) := [a_1^T;a_2^T;\dots;a_i^T]^T$. The disturbance variables at time $k$ are therefore $w_k = col(w_k^O, w_k^D)$, where $w_k^O = col(w_k^{O,1}, w_k^{O, 2}, \dots, w_k^{O, R})$, and $w_k^D = col(w_k^{D,1}, w_k^{D, 2}, \dots, w_k^{D, R})$. $w_k^{O, I}$ and $w_k^{D, I}$ are the number of passenger requests starting from and ending in Region $I$, respectively.
The output variable is $y_k = col(y^1_k, y^2_k, \dots, y^{R}_k) $, where $y^{I}_k$ states how many requests starting from Region $I$ are successful answered at time $k$. 

\subsection{Preliminaries}
A conventional approach to solving this fleet operation problem initially involves constructing a parametric model. For example, the AMoD dynamical system is approximated as a discrete-time linear time-invariant model in \cite{Andrea2019}, i.e.,
\begin{equation}
\Bigl\{
    \begin{array}{cc}
       x_{k + 1}&=Ax_k+Bu_k + B_d w_k\\
       y_k &=Cx_k+Du_k + D_d w_k,
       \end{array}\label{LTI_dist}
\end{equation}
where $x_k \in \mathbb{R}^{n}$ is the state vector. $u_k \in \mathbb{R}^{m}$ is the control input vector, $y_k \in \mathbb{R}^p$ is the output vector of the system at time $k \in \mathbb{Z}$, and $w_k \in \mathbb{R}^q$ is the disturbance variable. 


Learning $A, B, C, D, B_d$, and $D_d$ through system identification or defining them based on expert knowledge can be a challenging task. Hence, when these system matrices are unknown, yet input, output, and disturbance variables are measurable and accessible, we employ a data-enabled predictive control algorithm. This strategy leverages data-constructed Hankel matrices in lieu of a system model to provide an optimal predictive control policy. In the ensuing discussion, we first recall the main result of \cite{WILLEMS2005} regarding the non-parametric system representation.



\textbf{\textit{Definition 1. Persistent excitation \cite{WILLEMS2005}}}: 

Let $L,T_d\in \mathbb{N}$. The signal $z_k\in \mathbb{R}^m, k = 1, 2, \dots, T_d$ is persistently exciting of order $L$ if the Hankel matrix
\begin{equation}
\mathcal{H}_L(z):= \left[
    \begin{array}{cccc}
          z_1 & z_2  & \dots & z_{T_d-L+1}\\
          z_2 & z_3  & \dots & z_{T_d-L+2}\\
         \vdots &\vdots &\ddots &\vdots \\
         z_L &z_{L+1} &\dots &z_{T_d}
    \end{array}\right]
\end{equation}
has full row rank.

Let $T_{ini}, N\in\mathbb{N}$, $T_{ini}, N > 0$ be the lengths of time horizon for initial condition estimation and future prediction, respectively. Let $u^d$, $w^d$ and $y^d$ be historical trajectories of length $T_d$. The Hankel matrices constructed by $u^d$, $w^d$ and $y^d$ are partitioned into two parts, where the superscript `$p$' stands for `past data' and `$f$' for `future data' as
\begin{equation}\label{hankel_u}\left[
    \begin{array}{cc}
           U^p\\
          U^f
    \end{array}\right]:=\mathcal{H}_{T_{ini}+N}(u^d), 
\end{equation}
\begin{equation}\label{hankel_w}\left[
    \begin{array}{cc}
           W^p\\
          W^f
    \end{array}\right]:=\mathcal{H}_{T_{ini}+N}(w^d), 
\end{equation}
\begin{equation}\label{hankel_y}\left[
    \begin{array}{cc}
           Y^p\\
          Y^f
    \end{array}\right]:=\mathcal{H}_{T_{ini}+N}(y^d),
\end{equation}
where $U^p\in \mathbb{R}^{mT_{ini}\times(T_d-T_{ini}-N+1)}$ and $U^f\in \mathbb{R}^{mN\times(T_d-T_{ini}-N+1)}$ represent the first $mT_{ini}$ and last $mN$ block rows of $\mathcal{H}_{T_{ini}+N}(u^d)$, respectively.
Similarly for $W^p$, $W^f$, $Y^p$ and $Y^f$.




\textbf{\textit{Lemma 1. \cite{Huang2022}}}: Assume that historical trajectories of input $u^d$ and external disturbance $w^d$ of length $T_d$ are measured such that $col(u^d, w^d)$ is persistently exciting of order $T_{ini} + N + n$.  The vector $col(u,w,y)$ is a future trajectory of system (\ref{LTI_dist}) with the initial condition $col(u_{ini}, w_{ini}, y_{ini})$, if and only if there exist $g\in \mathbb{R}^{T_d-T_{ini}-N+1}$ such that 
\begin{equation}\left[
    \begin{array}{cccc}
         U^p\\
         W^p\\
         Y^p\\
         U^f\\
         W^f\\
         Y^f\\
    \end{array}\right] g = 
\left[
    \begin{array}{cccc}
         u_{ini}\\
         w_{ini}\\
         y_{ini}\\
         u\\
         w\\
         y\\
    \end{array}\right].\label{Hankel_uwy}
\end{equation}

This lemma indicates that the subspace spanned by the columns of the Hankel matrix, i.e., $col(U^P, W^p, Y^p, U^f, W^f, Y^f)$, corresponds to the subspace of all possible trajectories of the system (\ref{LTI_dist})\cite{Wang2023}. Thus, the Hankel matrix, built from raw data, can serve as a non-parametric model representing system behavior, thereby circumventing the system model's parameter identification. Next, we will employ optimal predictive control, utilizing the constructed Hankel matrices as substitutes for the system model to predict the future trajectory.

\subsection{Data-enabled predictive control for inter-regional vehicle repositioning}
Using the Hankel matrices to replace the system model, DeePC \cite{Coulson2018} solves an optimization problem in a receding horizon manner, attempting to obtain optimal control actions meanwhile satisfying input and output constraints.


 Let $N \in \mathcal{N}, N > 0$ be a given prediction horizon. At time $k$, we define $u_{ini} = col(u_{k - T_{ini}}, u_{k - T_{ini} + 1}, \dots, u_{k - 1})$. Similarly, $w_{ini} = col(w_{k - T_{ini}}, w_{k - T_{ini} + 1}, \dots, w_{k - 1})$, and $y_{ini} = col(y_{k - T_{ini}}, y_{k - T_{ini} + 1}, \dots, y_{k - 1})$. We formulate the optimization problem as follows\footnote{For simplicity, here and in \cref{control_obj}, we drop the dependency on $k$ in vectors collecting multiple samples of signals.}

\begin{equation}\label{deepc}
    \begin{array}{cc}
                 \displaystyle{\min_{g,u,y,\delta_y} f(u,y) + \lambda_g \Vert g\Vert_2 ^2 + \lambda_y \Vert \delta_y \Vert_2^2}\\
         subject \quad to \left[

 \begin{array}{cccc}
         U^p\\
         W^p\\
         Y^p\\
         U^f\\
         W^f\\
         Y^f\\

    \end{array}\right] g = 
\left[
    \begin{array}{cccc}
         u_{ini}\\
         w_{ini}\\
         y_{ini}\\
         u\\
         w\\
         y\\
    \end{array}\right]+ 
    \left[
    \begin{array}{cccc}
         0\\
         0\\
         \delta_y\\
         0\\
         0\\
         0\\
    \end{array}\right],\\

    u_{k + i}\in \mathcal{U},\forall i \in{0,1,\dots,N-1},\\
    y_{k + i}\in \mathcal{Y}, \forall i \in{0,1,\dots,N-1},
    \end{array}
\end{equation}
where $f(u,y)$ is the control objective function. $\Vert \cdot \Vert_{a}$ denotes the $a$-norm. The input and output constraints are $\mathcal{U}\subseteq \mathbb{R}^{mN}$ and $\mathcal{Y}\subseteq \mathbb{R}^{pN}$, respectively. 
The scalar $\lambda_g >0$ is a regularization parameter to avoid data overfitting.
Compared to (\ref{Hankel_uwy}), $\delta_y \in \mathbb{R}^{T_{ini}p}$ is added to the right side as an auxiliary slack variable. Moreover, a regularization $ \lambda_y \Vert \delta_y \Vert_2^2$ is incorporated in the objective to ensure the feasibility of the constraints at all times (see \cite{Coulson2018,Coulson2019}), where $\lambda_y >0$ is a weight coefficient. 

The vector $col(u_{ini}, w_{ini}, y_{ini})$ represents the most recent input, disturbance and output measurement.
Note that $u$ and $y$ are not independent of $g$ since by \cref{Hankel_uwy}, $u = U^fg$ and $y = Y^fg$. Solving (\ref{deepc}) gives an optimal control action sequence $u^* = col(u^*_k, u^*_{k+1}, \dots, u^*_{k + N - 1})$, and we only apply the first control input, i.e., $u^*_k$.

We define the economic objective as
\begin{equation}\label{control_obj}
    f(u, y) = \sum_{i =0 }^{N-1}- ||\mathcal{\alpha \cdot Q} y_{k + i}||_1 + ||\mathcal{R}u_{k+i}||_1,
\end{equation}
where $\mathcal{Q} \in \mathbb{R}^{1\times p}, \mathcal{R} \in \mathbb{R}^{1 \times m}$, are weight matrices for output and input, respectively. The first term is designed to encourage answering more requests. Moreover, the second term considers the rebalancing cost of relocating vehicles from one region to another. A weight parameter $\alpha >0$ balances the trade-off between these two factors. Its effect will be discussed in \cref{sec:alpha}.


In the context of AMoD systems, we can collect the historical demand requests $W^p$ and $W^f$, and timely measure the most recent demand $w_{ini}$. In addition, in this section, we assume that future requests can be predicted accurately (for example, see \cite{Ramon2017} for demand forecasting using Long Short-Term Memory neural networks). Thus, we utilize the true values for future demand $\tilde w$ to approximate the ideal, i.e., $w = \Tilde w$. Therefore, we add to (\ref{deepc}) the equality constraint

\begin{equation}\label{equ_w}
    W^fg = \Tilde w.
\end{equation}
More practical settings with inaccurate demand prediction will be discussed in \cref{noisydemand}.

Variable $e^I_k$ measures the number of empty vehicles in Region $I$ at time step $k$. Because we can only operate the unoccupied vehicles for rebalancing, given the availability of vehicles $e^I_k$, the input constraint should limit the number of vehicles that can be relocated at the current time step as follows

  \begin{equation}\label{inputconst}
         \sum_{J = 1}^{R} u^{IJ}_k = e^I_k,\quad \forall I \in{0,1,\dots,R}.
  \end{equation}

Both input and output variables are also required to be positive which leads to the following constraints defining the sets $\mathcal{U}$ and $\mathcal{Y}$ in (\ref{deepc}) 
    
\begin{equation}\label{inandoutpos}
    \begin{aligned}
u_{k + i}^{IJ} \geq 0&, \\
y_{k + i}^{I} \geq 0&,\\
\forall i \in{0,1,\dots,N-1}&. \quad \forall I, J \in {1,2,\dots,R}.
\end{aligned}
\end{equation}

\subsection{Implementation of upper-layer control actions}\label{subsec: deepc_implement}
Solving the optimization problem (\ref{deepc}) provides an optimal control sequence by $u^* = U^fg^*\in \mathbb{R}^{mN}$, therefore the elements of $u^*$ are not necessarily integer numbers. 
For implementation, $\lfloor u_k^{IJ} \rfloor$ empty vehicles in Region $I$ should be relocated to Region $J$ when $I \neq J$, where $\lfloor a \rfloor$ denotes the largest integer not greater than $a$; then the rest of the empty vehicles will stay in their current region. Moreover, let $\theta_k^{IJ}\in [0, 1]$ represent the empty vehicle flow transfer ratio from Region $I$ to Region $J$ as
\begin{equation}\label{theta}
    \theta_k^{IJ} = \frac{u_k^{IJ}}{\sum_{J = 1}^{R}u_k^{IJ}}.
\end{equation}
After applying the optimal control input $u_k$ given by the upper-level controller at time step $k$, when $k\Delta T < t < (k+1)\Delta T$, whenever a vehicle in Region $I$ becomes empty, it has a probability of $\theta_k^{IJ}$ of being relocated to Region J. This probabilistic approach ensures that the relocation of empty vehicles aligns with the optimal control strategy determined at each time step.
The procedure of implementing DeePC is summarized in \cref{alg:deepc}.
\begin{algorithm}
\caption{Upper-layer Control: DeePC Algorithm for Inter-regional Vehicle Transfer Guidance}\label{alg:deepc}
\textbf{Input:} The Hankel matrices $U^p, U^f, W^p, W^f, Y^p, Y^f$ constructed from the historical data $u^d, w^d$ and $y^d$. \\ 
\textbf{Data Measurement:} At time $k$, collect the most recent measured data $u_{ini}, w_{ini}$ and $y_{ini}$ of length $T_{ini}$, the number of current empty vehicles in each region.\\
\textbf{Output:} Optimal control input $u_k$, transfer ratio $\theta_k^{IJ}$. \\
\begin{algorithmic}
\State 1) At time $k$, obtain $g^*$ by solving the optimization problem (\ref{deepc}),
\State 2) Compute the optimal control input sequence $u^* = U^fg^*$, where $u^* = col(u^*_k, u^*_{k + 1},\dots,u^*_{k+N-1})$,
\State 3) Apply the control input, i.e., $u_k = \lfloor{u^*_k}\rfloor$, calculate $\theta_k^{IJ}$ by \cref{theta},
\State 4) $k \leftarrow k+1$, return to Step 1).
\end{algorithmic}
\end{algorithm}

\section{Coverage-based lower-layer control}\label{sec: lower}
In the lower-level control, we address the management of empty vehicles that are instructed by the upper-layer controller to remain in their current regions. Specifically, in this work, we implement a coverage control-based algorithm to provide intra-regional position guidance, building upon the methodology proposed in our earlier work \cite{Pengbo2023a}.  

\begin{figure*}[htb]
\centering
\begin{subfigure}[ht]{0.3\textwidth}
\includegraphics[width=\textwidth]{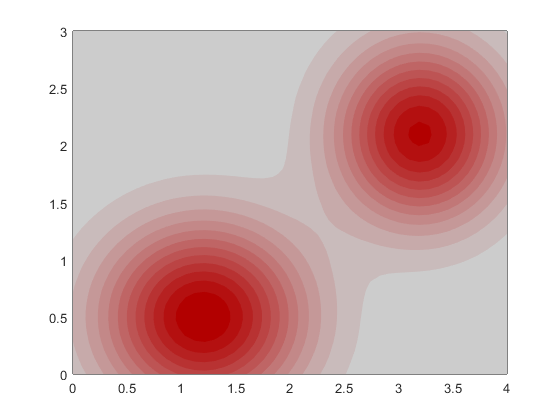}%
 \caption{The Density distribution $\phi$}\label{fig:density}
\end{subfigure}
\begin{subfigure}[ht]{0.3\textwidth}
\includegraphics[width=\textwidth]{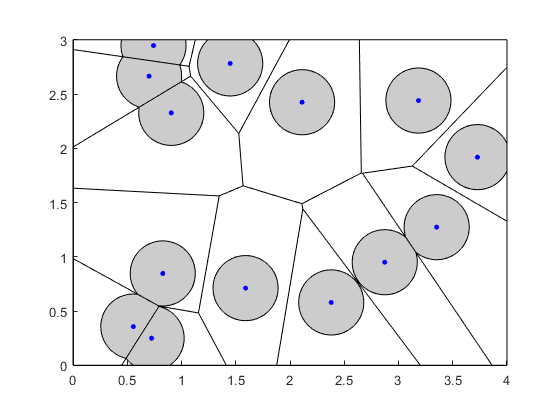}%
 \caption{Initial configuration of agents}
\label{fig:init}
\end{subfigure}
\begin{subfigure}[ht]{0.3\textwidth}
\includegraphics[width=\textwidth]{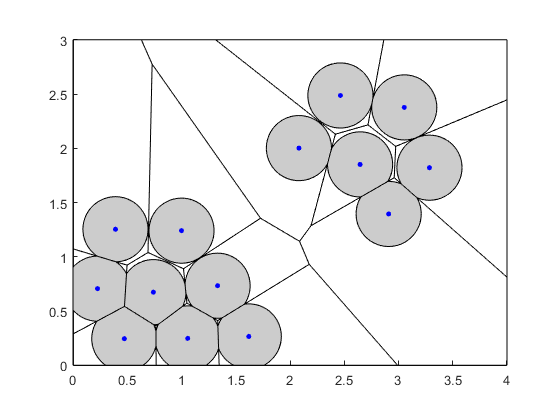}%
 \caption{Optimal configuration of agents}
\label{fig:converge}
\end{subfigure}
\caption{Coverage control algorithm has been applied in mobile multi-agent problems.}
\label{fig:concept_coverage}
\end{figure*}

\cref{fig:concept_coverage} shows a multi-agent system where the coverage control algorithm is in action. Each agent is responsible for executing a specific activity/task within a limited radius $r$ (i.e. $r$-limited Voronoi cell \cite{Sotiris2018A}, shown as the gray disks in \cref{fig:init} and \cref{fig:converge}), such as wildfire monitoring or underwater exploration \cite{Jorge2004}. The occurrence of these activities follows the distribution $\phi$, as shown in \cref{fig:density}. The coverage objective is to minimize the expected distance between a task location and the agent that will accomplish it. By directing the agents at the centroid of their covered area, starting from an initial random configuration shown in \cref{fig:init}, the coverage control algorithm guides the vehicles to converge to an optimal configuration, as depicted in \cref{fig:converge}. 
Motivated by the ability of coverage control algorithms to effectively coordinate and control mobile agents, we apply it to guide empty vehicles in order to align their configuration with passenger demand density.

\subsection{Coverage control for intra-regional position guidance}\label{subsec:formula_lower}

Consider the structure of a city road network, the intersections can be extracted as nodes and road segments as links. We assume one vehicle can serve passengers within a limited range of $r$. Instead of using a continuous space coverage approach as shown in \cref{fig:concept_coverage}, we consider the `covered area' of one vehicle as defined by the Voronoi partition on the graph \cite{Martin2000}. The nodes within the $r$-limited graph Voronoi cell of vehicle $i$ are characterized by two properties: 1) They are located within an $r$-limit distance from the vehicle's current position $p_i$, and 2) they are closer to vehicle $i$ than to any other vehicles.

\begin{figure}[htb]
     \centering
         \includegraphics[width=0.45\textwidth]{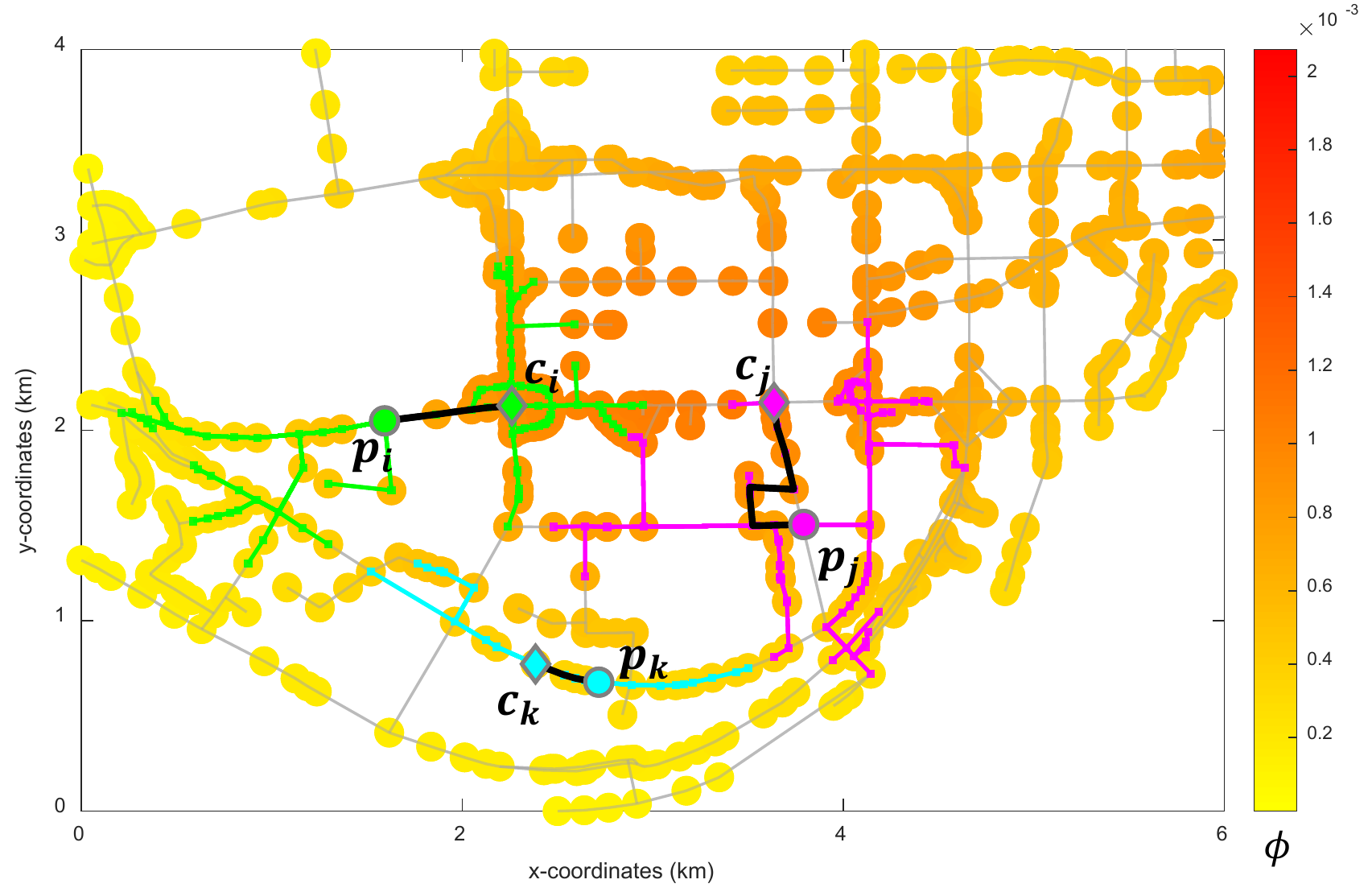}
         \caption{Illustration of Voronoi partition on the graph. Here we show a 3-vehicle case. Vehicle $i$ is located at $p_i$ (green round marker), its covered nodes are shown as green-filled squares surrounding it, and are connected by lines to show the `covered area' of vehicle $i$. The centroid of its corresponding $r$-limited Voronoi Cell is denoted as $c_i$, shown as a green diamond marker. 
         For each vehicle, the shortest path from the current position to its centroid is illustrated by the black trajectory. A heat map is superimposed over the road network representing the demand distribution $\phi$. The color gradient ranges from yellow to red, with yellow signifying lower demand probabilities and shades of red indicating increasingly higher demand levels.}\hfill
         \label{fig:graphVoro}
     \end{figure}
     
The centroid of the $r$-limited graph Voronoi cell can be computed using an integer optimization problem (see \cite{Durham2012}), which minimizes the expected distance between a passenger request and the vehicle. The optimal position configuration for all vehicles can be achieved by locating each vehicle at the centroid of its respective Voronoi cell, as suggested by \cite{Joseph2011}. 

The demand distribution can be estimated from historical taxi trips. In \cref{fig:graphVoro}, it shows that the passenger demand is unevenly distributed over different districts even within a single region. By directing the idle vehicles to move toward their centroids, this coverage control algorithm guides vehicles
to gather around the high-demand districts, which can be observed that vehicles are moving toward areas with a more intense red shade in \cref{fig:graphVoro}.


The implementation of this lower-layer controller is described in Algorithm \ref{alg:lower}. This algorithm is designed in a distributed manner \cite{Pengbo2023a}, and can be implemented on each vehicle to generate its own position guidance. Consequently, it is computationally efficient and suitable for large-scale systems. Moreover, compared with the regional-level instruction given by the upper-layer control which indicates `which region to go', this lower-layer control provides finer-grained guidance, specifying `which intersection to go'. In addition, at the end of each upper-level time step, aggregated local data, such as the number of available vehicles in each region and the number of completed requests in each region, are sent back to the upper layer. This information is then used to determine the next upper-level command.

\begin{algorithm}[ht]
\caption{Lower-layer Control: Coverage Control Algorithm for Intra-regional Position Guidance}
\label{alg:lower}
At each time step $k'$:\\
\For{each idle vehicle $i$}{
    \textbf{Communication:} Collect the current coordinates of the idle vehicle and its neighbor vehicles. \\
    \textbf{Voronoi Cell Computation:} Compute the Voronoi cell on the graph based on the collected coordinates. \\
    \textbf{Centroid Computation:} Calculate the centroid of the Voronoi cell on the graph. \\
    \textbf{Movement:} Set the calculated centroid as the rebalancing destination for this idle vehicle and move towards it by following the shortest path calculated by Floyd-Warshall Algorithm \cite{Floyd1962}.
}
\end{algorithm}

\subsection{Hierarchical Structure}
The hierarchical control framework integrates the upper and lower-layer control algorithms previously described in \cref{subsec: deepc_implement} and \cref{subsec:formula_lower}. The upper-layer controller provides inter-regional vehicle transfer instructions, while the lower-layer controller guides the intra-regional vehicle repositioning. This proposed framework is detailed in Algorithm \ref{alg:hier}.

\begin{algorithm}[ht]
\caption{Hierarchical Structure for Empty Vehicle Repositioning in AMoD Systems}
\label{alg:hier}
At each time step $k$:\\
\textbf{Input:} Online measurement of input-output data, and external disturbance as introduced in \cref{subsec:motivation_upper}.\\
\textbf{Compute:} Solve the optimization problem defined in (\ref{deepc}). Refer to the upper-layer controller described in Algorithm \ref{alg:deepc}.\\
\textbf{Output:} Determine the number of empty vehicles that should either relocate to other regions or stay in their current regions.\\
At each time step $k'$:\\
\For{each empty vehicle that is relocated to another region}{
    \textbf{Compute:} Select the nearest node in the desired region as the relocation destination\footnote{}.\\
    \textbf{Move:} Move towards the selected destination node by following the shortest path computed using the Floyd-Warshall Algorithm \cite{Floyd1962}.}

\For{each empty vehicle that should stay in its current region}{
    $\rightarrow$ Refer to the lower-layer controller described in Algorithm \ref{alg:lower} for further position instruction.\\
}

\end{algorithm}

\footnotetext{ Sending vehicles to `the nearest road intersection in their designated region' aims to complete inter-regional relocations as quickly as possible, as vehicles will follow the shortest path so that they will arrive in their designated region swiftly. Future work will investigate how to select the relocating destinations.}



\section{Case Study}\label{sec: results}

The proposed method is tested on an AMoD simulator (see \cite{Caio2021}) representing the urban road network of Luohu and Futian districts in Shenzhen, China. The network consists of $1858$ intersections and $2013$ road links.

\begin{figure}[htb]
    \centering
    \begin{subfigure}{0.48\textwidth}
        \includegraphics[width=\textwidth]{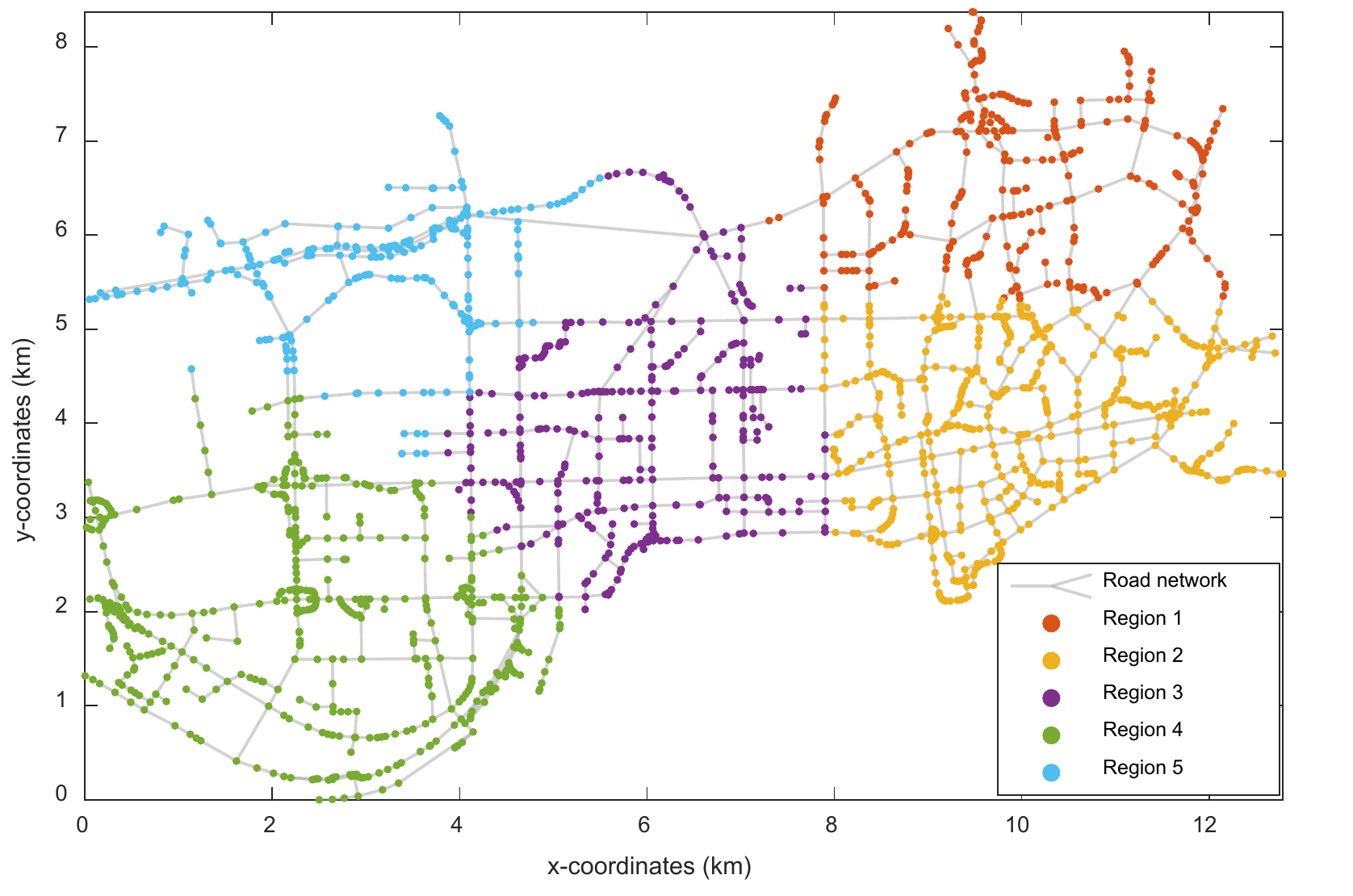}
        \caption{The city area is clustered into 5 non-overlapped regions.}
        \label{fig:kmean5}
    \end{subfigure}\\
    \centering
    \begin{subfigure}[b]{0.48\textwidth}
        \includegraphics[width=\textwidth]{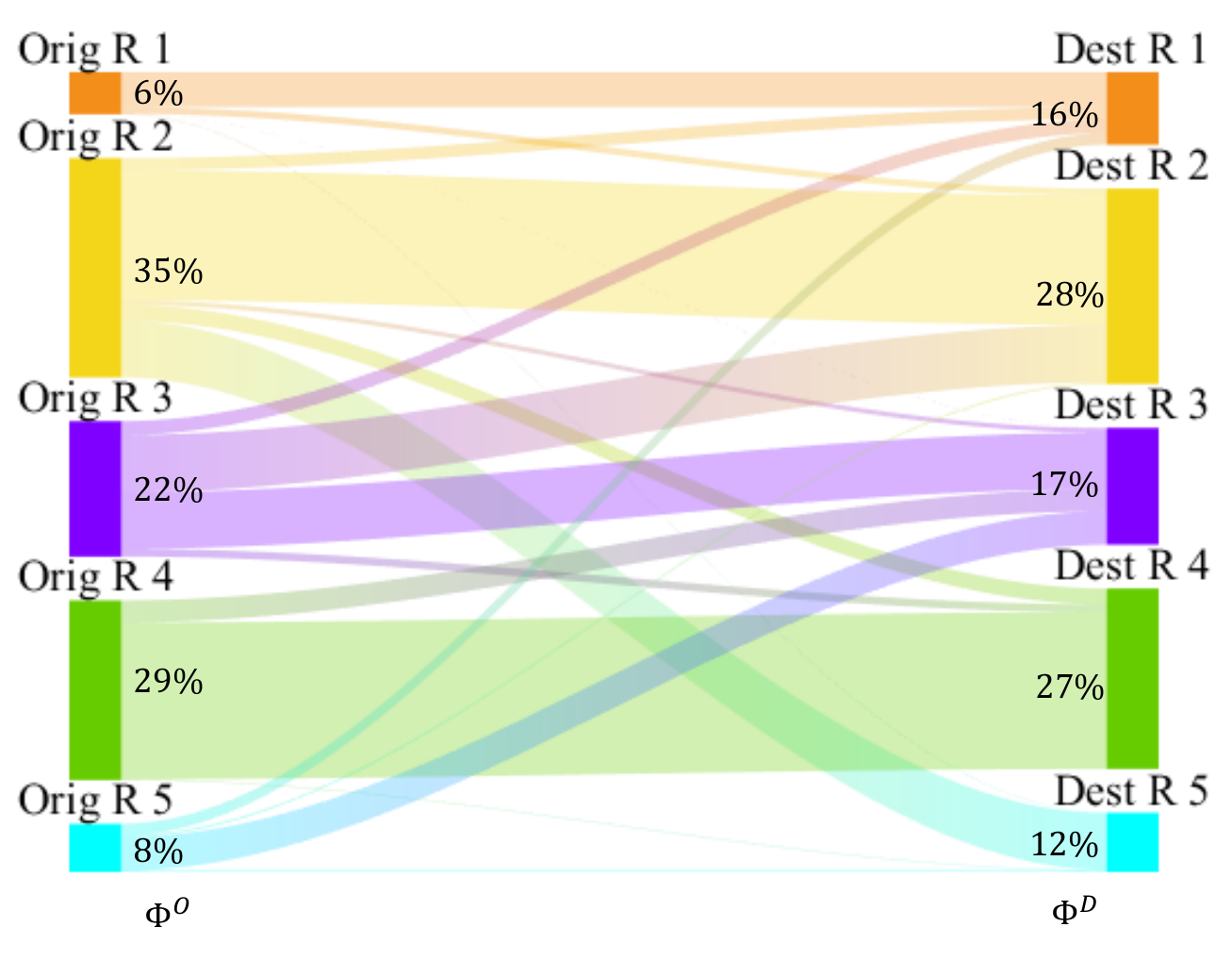}
        \caption{A Sankey diagram showing the trip distribution from Region $I$ to Region $J$}
        \label{fig:sankey}
    \end{subfigure}
    \caption{City partition and trip origin and destination distribution.}
    \label{fig:cluster5}
  \vspace{-0.3cm}
\end{figure}

The experimental scenario focuses on a significant discrepancy between the trip origin and destination distributions, which is described in detail in \cite{Pengbo2022}. The region of interest is clustered into $R$ regions using $K$-means clustering as it shows in \cref{fig:kmean5}, where $R$ is set to 5 in this study. From the historical taxi trips, we can obtain the trip origin distribution (i.e., regional demand probability) for these 5 regions $\Phi^{O} = [0.06, 0.35, 0.22, 0.29, 0.08 ]$. Similarly, the trip destination distribution is $\Phi^D = [0.16, 0.28, 0.17, 0.27, 0.12]$. \cref{fig:sankey} represents a Sankey flow diagram, which illustrates the trip origin and destination distributions. The thickness of the lines would correlate to the volume of trips. It reveals that Regions 2,3, and 4 have higher demand than Regions 1 (6\%) and 5 (8\%), however, more trips terminate in Regions 1 (16\%) and 5 (12\%).

\noindent

\subsection{Data collection procedure for upper-layer control}

To facilitate the implementation of the upper-layer DeePC-based controller, the historical input $u$ and disturbance $w$ trajectories should be collected for building the Hankel matrix.
As stated in (\ref{Hankel_uwy}), the historical data $col(u^d, w^d)$ must be persistently exciting. To satisfy this requirement, on the one hand, the control actions, in terms of the vehicle flow transfer ratio as described in (\ref{theta}), were generated randomly in the data collection procedure. On the other hand, the issued time of passenger requests was sampled from a constant rate Poisson distribution. To obtain data that accurately reflects the system dynamics when both layers are functional, note that the lower-layer coverage control is activated during data collection.

Simultaneously, the number of answered requests (i.e., the output) was measured as well. The aforementioned data were further utilized to construct the Hankel matrices as described in (\ref{hankel_u}), (\ref{hankel_w}), and (\ref{hankel_y}), which had a length of $T_d = 3000$ with a sampling interval of $T_u = 10$ minutes. The values of all hyperparameters used in this case study are provided in \cref{tab:hyper}.

Meanwhile, considering the design of the control objective in (\ref{control_obj}), we measured the rebalancing trip lengths between regions to form the penalty weights for control inputs, in particular, $\mathcal{R} = \{ L_{IJ}\}$, where each element represents the average rebalancing trip length (unit: km) from Region $I$ to Region $J$ (listed in \cref{tab:hyper}). The upper controller only takes the inter-regional repositioning cost into account, therefore diagonal elements are set to 0, indicating that vehicles staying within their current regions do not cost extra fuel.
And the weight for output is devised as $\mathcal{Q} =
\Phi^O$, where higher weights are assigned to regions with higher demand in accordance with the demand probability $\Phi^O$. 
The hyperparameter $\alpha$ is chosen at 150, which provides good performance balancing the objectives of maximizing fulfilled requests and minimizing rebalancing costs.
Further details on the tuning process will be provided in \cref{sec:alpha}.

\subsection{Simulation Results}

\begin{table*}[ht]
  \begin{center}
    \caption{Performance metrics (30s)}\label{table_hier}
     \vspace{2ex}
    \begin{tabular}{c||c|c|c|c} 
    \hline
       \textbf{   }&\textbf{answer rate ($\%$)} & \textbf{waiting time ($s$)} & \textbf{rebalancing distance ($km$)} & \textbf{VUR} ($\%$)\\
             \hline
      Upper + Lower &87.6   &133.9 &3658.0   & 87.1  \\
      \hline
      Upper Only  &85.4 &139.4   &2944.2    &86.3\\
      \hline
      Lower Only &81.2   & 142.5&  2547.6 &  83.8\\

      \hline
      No Control   &51.6 & 156.9  &0.0      & 55.6 \\

      \hline \hline
                 LP  \cite{AlonsoMora2017A}     &74.7 & 151.6 & 1337.9 & 77.6\\
      \hline


      No Control (fleet size = 450)   &69.2 & 140.4   &0.0      & 47.0 \\
      \hline
    \end{tabular}
  \end{center}
\end{table*}
To create a better interaction between the two layers and also have a larger number of vehicles available to offer service even during relocation, repositioning vehicles are permitted to pick up passengers and are thus classified as `answer available' vehicles. On the passenger side, every issued request will remain in the matching pool for $T_m$. 
If the nearest empty vehicle can reach the passenger within a predetermined time threshold $T_w$ after the request has been issued, the vehicle and passenger will be matched. Once matched, cancellations are not permitted in this study. Conversely, if no available vehicles are found within the $T_m$ timeframe, the passenger will not wait anymore and cancel this request. Additional details about the parameters characterizing the simulation scenarios and the control scheme are given in \cref{tab:hyper}.

We conduct a comparison using four performance metrics in this study. Firstly, the \textbf{answer rate} is evaluated as the ratio of successfully answered requests to the total number of requests. Secondly, the \textbf{average waiting time} is computed by dividing the total waiting time for passengers, from the moment their requests are issued until they are picked up by vehicles, by the number of all answered requests. Moreover, the \textbf{rebalancing distance} is measured as the cumulative distance traveled due to both inter- and intra-regional repositioning of the empty vehicles. Finally, the \textbf{Vehicle Utilization Rate (VUR)} captures the proportion of time vehicles are engaged in serving passengers compared to their idle time. It ranges from $0$ to $1$, with higher values indicating a higher proportion of time spent serving passengers and lower idle times, thus demonstrating a more efficient utilization of the vehicles.






In this section, we evaluate the following control strategies: (i) the proposed hierarchical control method, referred to as the \textit{Upper + Lower}, which combines both the high-level and low-level controllers; (ii) using only the high-level controller (DeePC-based) to instruct inter-region vehicle transfer. This strategy is denoted as \textit{Upper Only}, where relocating vehicles target the nearest road intersection in their designated new region as their relocation destination, and upon reaching, they are not further guided to any specific intersections. For the sake of simplicity and generality, the inter-regional relocating vehicles are selected randomly from the available fleet in each region for the aforementioned two strategies. (iii) using only the low-level controller (coverage control-based) to provide intra-regional position guidance commands, where empty vehicles do not relocate among different regions, referred to as \textit{Lower Only}; and (iv) a scenario with no active controllers, where empty vehicles remain at their current positions unless serving a passenger, referred to as \textit{No Control}. Moreover, we benchmark our approach against a fleet rebalancing strategy referred to as \textit{LP} in \cref{table_hier}, which reallocates idle vehicles to intersections with pending/unassigned requests \cite{AlonsoMora2017A}. This strategy is implemented through integer linear programming to minimize total travel time between vehicle-request pairs, subject to the condition that either all pending requests or all idle vehicles are allocated. Compared with our proposed framework which combines centralized (upper-layer) and distributed (lower-layer) ways to control the fleet, `LP' policy is designed in a centralized manner and requires real-time access to information about unfulfilled requests. We test the performance of `LP' policy when it is operated every $30 s$ as implemented in \cite{AlonsoMora2017A}.

One 3-hour experiment is carried out with a fleet size of 300, and around 5500 requests are issued in total. Without loss of generality, the initial distribution of vehicles is sampled randomly but uniformly across the urban area. All policies are tested with the same initial distribution of vehicles and the same passenger request profile. As indicated in \cref{table_hier}, the `Lower Only' policy shows significant advantages over the `No Control' policy, fulfilling more requests and reducing waiting times.  `Upper Only', operating at a regional level and dispatching vehicles to move across regions, therefore shows a higher rebalancing distance than `Lower Only'. Nonetheless, its control objective design, as per (\ref{control_obj}), leads to a better answer rate compared to `Lower Only'. Integrating the lower-layer inter-regional rebalancing to `Upper Only', our proposed hierarchical framework increases the rebalancing distance as expected, while it also notably enhances over answer rate, average waiting time, and VUR, compared to when the upper or lower-layer controller is applied in isolation. Compared with our method, the `LP' strategy focuses solely on current unassigned requests, which does not relocate a sufficient number of empty vehicles. This results in reduced rebalancing distances, but shows limited improvements over `No Control' in both answer rate and average waiting time.

Our study also evaluated the `No Control' policy using larger fleets. Notably, our framework outperforms the `No Control', even the latter one is operated with $50\%$ more fleets (fleet size = 450). However, such an increase may inflate the operational costs of AMoD companies and worsen network congestion, as it has a lower VUR than the case with fleet size = 300. This highlights our method's potential for practical application, as it improves performance while facilitating smaller, cost-effective fleets. 



\begin{figure}[htb]
    \centering
    \includegraphics[width=0.5\textwidth]{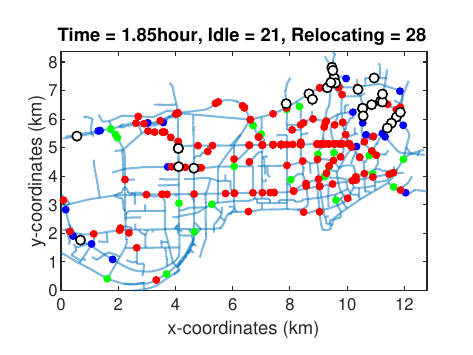}
    \caption{A snapshot of the network with ride-hailing fleet: relocating vehicles (black circles, operated by the upper-layer controller), idle vehicles (blue dots, operated by the low-layer controller), occupied vehicles traveling to pick up (green) and drop off passengers (red). 
The demo video of the proposed method is available on Youtube: \textcolor{blue}{\url{https://youtu.be/x2UnJLx_wSM}} and the demo video of `No Control' policy can be found via \textcolor{blue}{\url{https://youtu.be/_zZEQEHn7vE}}}\label{fig:snapshot}

\end{figure}

\cref{fig:snapshot} shows a snapshot of the AMoD simulator. Due to the discrepancy between trip origin and destination distributions, more empty vehicles can be observed in Regions 1 and 5, revealing the vehicle oversupply in these areas. Thanks to the upper-layer controller, a significant portion of these vehicles is then directed to other regions with high demand, resulting in the presence of black circles (representing relocating vehicles) in those regions. In contrast, empty vehicles in Regions 2 and 4, indicated by blue dots, are instructed by the lower-layer controller to remain in their current regions since they are already located in advantageous positions. The demo video further showcases the inter-region movements of these blue dots as facilitated by the lower-layer controller. Both the upper and lower-layer commands are aligned with the passenger demand distribution, enhancing the service level of the AMoD system. 

\begin{figure*}[htb]
    \centering
    \includegraphics[width=1\textwidth]{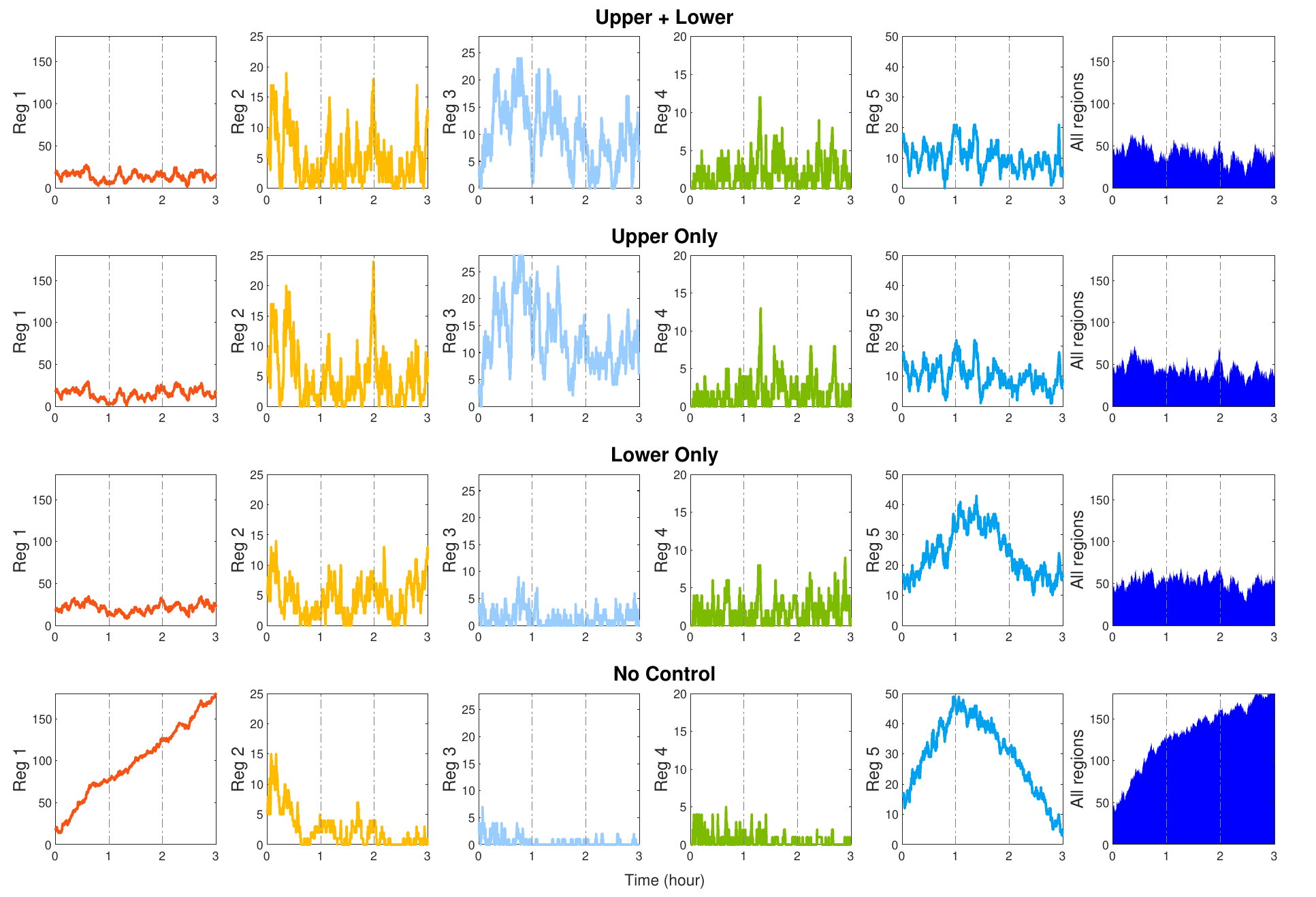}
    \caption{The evolution of the number of empty vehicles under different control methods. The first five columns on the left depict how the number of empty vehicles changes over time in each region, while the final column shows the total count of empty vehicles in all regions.}\label{fig:evolution}
\end{figure*}

\cref{fig:evolution} depicts the evolution of the number of empty vehicles in each region, along with the aggregate count of all empty vehicles, under different repositioning policies. Under `No Control' policy, where no intervention is implemented, an increasing number of empty vehicles accumulates in Regions 1 over time, which can be seen in the demo video and in \cref{fig:snapshot} as well. This configuration is unfavorable as these vehicles are waiting for longer times before serving any passengers, resulting in a large number of request cancellations and the lowest VUR in \cref{table_hier}. Moreover, it depletes empty vehicles in Regions 2, 3 and 4, which indicates the undersupply in high-demand regions. The `Lower Only' policy partially mitigates this issue by directing vehicles towards high-demand areas within their current region, as evidenced by the reduced number of empty vehicles in Regions 1 and 5 compared with `No Control' policy. However, this approach lacks coordination between different regions, leading to suboptimal results such as an oversupply in Region 5.
The `Upper Only' method addresses this problem by continuously balancing the vehicles from low-demand regions to high ones. Therefore, there are more available vehicles in Regions 2 and 3. The proposed full framework demonstrates superior performance in terms of VUR that throughout the simulation there are only a few vehicles empty. Notably, there is a larger proportion of empty vehicles in high-demand regions compared with other policies, indicating improved vehicle accessibility for passengers and reduced waiting times. 

The results highlight the complementary nature of the different layers in the framework. The upper layer, based on predictive control, takes an aggregated view of the traffic system and adjusts fleet distribution among regions proactively. However, it lacks the precision required for microscopic vehicle guidance within each region. In contrast, the lower layer employs a coverage control-based mechanism to guide idle vehicles at the intersection/node level, while it lacks coordination across different regions. Therefore, the combination of the layers in a hierarchical control structure leverages the strengths of both macroscopic and microscopic control approaches, resulting in improved system performance.

\subsection{Sensitivity to $\alpha$}\label{sec:alpha}


During the design of the upper-layer controller, an economic objective is considered, see (\ref{control_obj}). This objective aims to strike a balance between maximizing the number of answered requests and minimizing the fuel cost associated with vehicle repositioning. The trade-off between these two factors is controlled by the parameter $\alpha$, which represents the relative weight assigned to each term.

\begin{figure}[htb]
    \centering
    \includegraphics[width=0.49\textwidth]{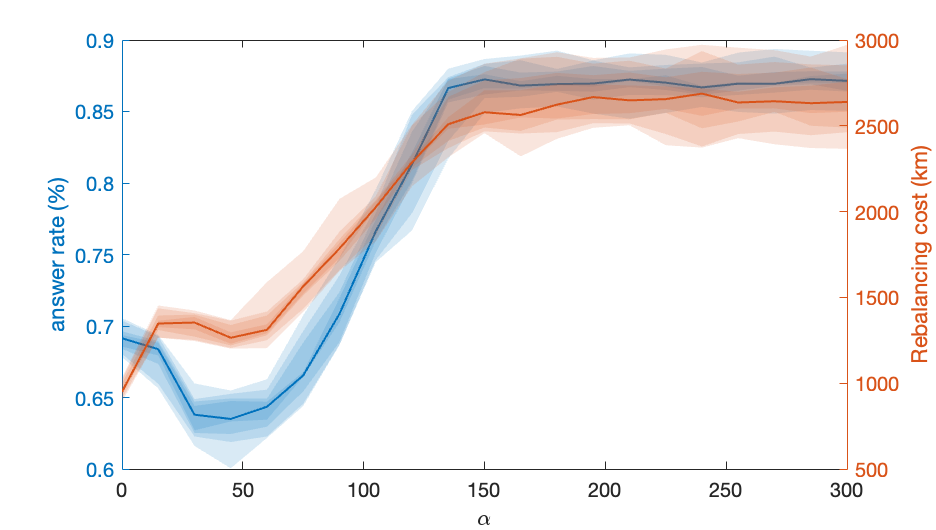}
    \caption{The effect of varying the parameter $\alpha$ on the answer rate (left y-axis) and the rebalancing cost (right y-axis). The solid lines represent the mean values for 15 runs, while the
shaded areas indicate the percentiles for varying degrees ($25\%$,
$50\%$, $75\%$, and $90\%$).
    }\label{fig:alpha}
\end{figure}

\cref{fig:alpha} illustrates the impact of different $\alpha$ values, ranging from $0$ to $300$, on the answer rate and accumulated rebalancing cost $\sum_k||\mathcal{R}u_k^*||$ in the second factor in (\ref{control_obj}). As $\alpha$ increases, the answer rate initially shows improvement, indicating a higher proportion of requests being successfully fulfilled. At the same time, the total rebalancing cost objective is increasing, implying that more vehicles are instructed to do an inter-regional transfer. The increase in answer rate diminishes after $\alpha \geq 150$ as visualized in \cref{fig:alpha}. 

\begin{figure}[tb]
    \centering
    \begin{subfigure}{0.44\textwidth}
        \includegraphics[width=\textwidth]{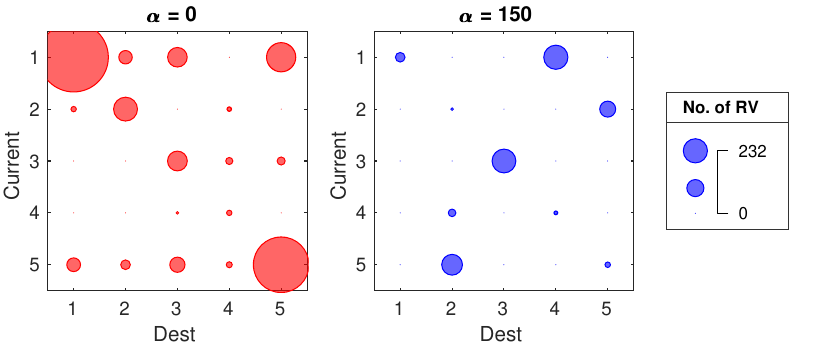}
    \caption{Bubble chart for total control commands $\sum u_k^*$. Each bubble is positioned on a grid that represents the `current region' (y-axis) and `destination region' (x-axis). The size of each bubble corresponds to the number of vehicles being relocated from one region to another. The larger the bubble is, the more vehicles are being relocated.}
    \label{fig:bubble}
    \end{subfigure}\\
    \centering
    \begin{subfigure}{0.35\textwidth}
        \includegraphics[width=\textwidth]{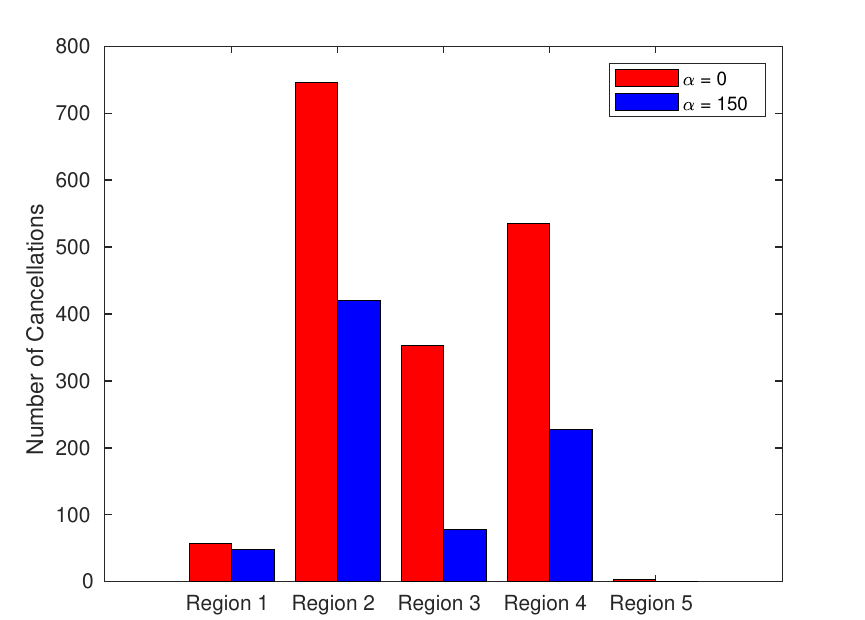}
        \caption{Bar chart for total cancellation in each region}
        \label{fig:cancellation}
    \end{subfigure}
    \caption{Comparison of $\alpha = 0$ and $\alpha = 150$. ‘Relocating Vehicles’ is abbreviated as `RV'.}
\end{figure}

The bubble diagrams \cref{fig:bubble} show a $5 \times 5$ matrix representing the accumulated control commands $\sum_k u_k^*$, illustrating an overview of how vehicle relocations are managed across the different regions of the city. It can be seen that, when $\alpha = 0$, the largest bubbles are located at the diagonal position, indicating that most vehicles are asked to stay at their current regions out of the sake to save the rebalancing cost; when $\alpha = 150$, more vehicles will move to other regions. 
In addition, the general bubble size for $\alpha = 0$ is bigger than $\alpha = 150$, implying the former case has many more vehicles unoccupied in low-demand regions and more cancellations shown in \cref{fig:cancellation}, which is similar to the case `No Control' shown in \cref{fig:evolution}.

It is interesting to see vehicles in Region 1 are heading to Region 4 when $\alpha = 150$, and vehicles in Region 5 are heading to Region 2. This can be caused by the geometric nature of this specific partition where the route traveling from Region 1 to Region 4 crosses Region 3. A similar observation applies when traveling from Region 5 to Region 2. 
In our test, one relocating vehicle is classified as `answer available', which means that it will terminate its rebalancing movement whenever it is matched with a passenger as the best candidate to answer this request. Our data-driven method successfully captures this complex behaviour and instructs vehicles to pass through more high-demand regions.

These results emphasize the importance of carefully considering the trade-off between these two objectives for the TNC company. A low answer rate can lead to dissatisfied customers who may change to alternative transportation modes. On the other hand, a large rebalancing distance signifies higher operational costs and increased maintenance efforts for the fleet. Finding an appropriate balance between these factors is crucial for achieving optimal performance and customer satisfaction.

In addition to $\alpha$, the tuning process for other hyperparameters, such as $\lambda_y$ and $\lambda_g$, involves defining a grid of possible values and evaluating the performance for each combination of hyperparameters.  In our case study, we selected the hyperparameter configuration that maximizes the answer rate.

\subsection{Noise in the Demand}\label{noisydemand}

In the previous discussions, we assumed perfect knowledge of the external disturbance $w_k$, as represented in \cref{equ_w}. However, in reality, the accuracy of demand predictions can be affected by various factors. In this section, we explore the impact of inaccurate demand predictions by introducing unbiased noise to the predicted demand, denoted as $\hat{w}_k$.

With a bit of abuse of notations, consider the ground truth demand $\Tilde{w}_k$ as a matrix, defined as $\Tilde{w}_k = \{\Tilde{w}_k^{IJ}\}$. $\Tilde{w}_k^{IJ}$ represents the element at $I$-th row and $J$-th column, indicating the number of requests starting from Region $I$ and ending in Region $J$ at time step $k$. Here, $I, J = \{1, 2, \dots, R\}$.

To account for random and recurrent variations in demand, we introduce unbiased noise to the predicted demand based on a Gaussian distribution (see \cite{Nikolas2013}), which reflects the inherent variability and uncertainty in the demand patterns. Specifically, the noisy predicted demand $\hat{w}_k^{IJ}$ is obtained as 
\begin{equation}
     \hat{w}_k^{IJ} = max(0, \quad \Tilde{w}_k^{IJ} + \mathcal{N}(0, \sigma^2)),
 \end{equation}
 where $\sigma^2$ represents the variance for the noise. 
Then we respectively compute the row-wise and column-wise sum of the matrix $\hat{w}_k$ to obtain vector $\hat{w}_k^O \in \mathbb{R}^{R} $ and $\hat{w}_k^D \in \mathbb{R}^{R}$.
Instead of assuming the ideal condition with perfect knowledge of the ground truth value $\Tilde{w}_k$ in (\ref{equ_w}), we test with different levels of noise in this section, i.e. $W^fg = col(\hat{w}_k^O, \hat{w}_k^D)$. Signal-to-Noise Ratio (SNR, unit: dB) is used to quantify how much the demand has been corrupted by noise, which is defined as the ratio of the sum of the variances of the demand ($var_{\Tilde{w}_k^{IJ}}$) to the variance of the noise, as follows: 
\begin{equation}
    SNR = 10 \times \log_{10}\left(\frac{\sum_{I=1}^R \sum_{J=1}^R var_{\Tilde{w}_k^{IJ}}}{\sigma^2}\right)
\end{equation}

\begin{figure}[ht]
    \centering
    \includegraphics[width=0.5\textwidth]{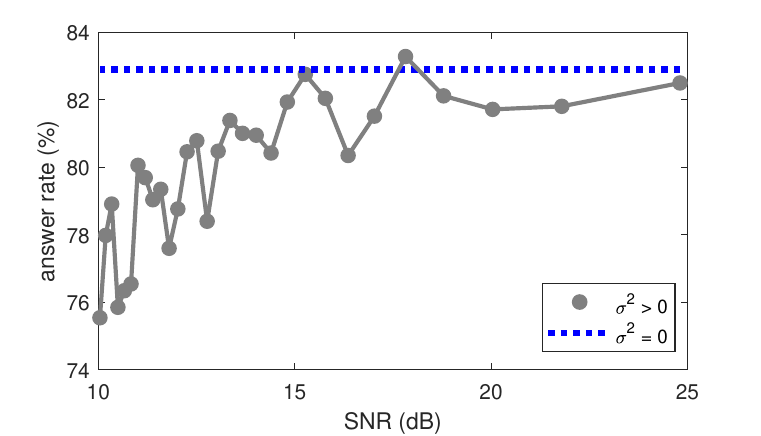}
    \caption{Scatterplot illustrating the relationship between answer rate and increasing noise variance ($\sigma^2$). The blue dotted line shows the performance under ideal conditions with perfect demand knowledge (i.e., when $\sigma^2 = 0$).
    }\label{fig:sigma}
\end{figure}

The impacts of demand noise on the answer rate are depicted in \cref{fig:sigma}, where the SNR varies between 10 and 25dB. As the noise level increases, a notable decline in the answer rate occurs. It suggests that inaccurate demand forecasts can lead to an incorrect system's nonparametric representation, resulting in suboptimal control commands. Despite being impacted by the noise, our proposed method still outperforms the `No Control' policy, which shows a performance of $51.6\%$ (see \cref{table_hier}).

\section{Conclusion}\label{sec: conclusion}
This paper has introduced a hierarchical control structure aimed at addressing the empty vehicle rebalancing challenge inherent in autonomous Mobility-on-Demand systems. The structure features two levels of control, each providing distinct advantages. At the upper level, our method avoids modeling the system. Instead, it leverages a Hankel matrix constructed from historical data to offer regional position guidance. The lower level introduces a coverage control-based approach that provides more precise position guidance for each individual vehicle within each region. The strength of our proposed structure lies in the efficient coordination it facilitates between the actions of the upper-level controllers, which manage aggregated traffic components, and the self-governance of individual vehicles at the lower level. Our proposed algorithm has been put to the test on a discrete city map using real road network geometry from the city of Shenzhen. The results have demonstrated that our method outperforms other approaches, achieving superior results in terms of answer rate, waiting time, and vehicle utility rate. Moreover, we have discussed how hyperparameters affect the system performance.

This study has presented a general and flexible structured strategy that permits modifications at each layer. This adaptability allows us to test alternative algorithms at each layer of this proposed structure, to further refine and optimize our solution for managing autonomous Mobility-on-Demand systems. One interesting future direction is to compare this data-driven method with model-based methods, where the model can be obtained via system identification or one existing dynamic model, e.g., the fluidic model in \cite{Pavone2011}. In addition to the unbiased noise in demand, the biased noise, e.g. such as unexpected adverse weather, can strongly affect the demand pattern. It is interesting to adapt our current method for time-varying cases, for example, online adaptation of the Hankel matrix.
Future research could also explore integrating monetary utilities to directly define the weight $\alpha$ which balances the answer requests and the rebalancing cost. In addition, the use of more advanced hyperparameter optimization methods, such as Bayesian optimization, can help fine-tune the values of $\lambda_g$, $\lambda_y$ more efficiently.

Currently, we use K-means clustering to partition the urban area, selecting the number of clusters based on the Elbow method \cite{Ketchen1996}. This choice of cluster count is also made with the consideration of balancing the trade-offs between region size and the efficiency of the hierarchical control structure. A higher number of partitions increases the computational cost of solving the upper-layer optimization problem, as the constraints and decision variables in the DeePC scale with the square of the number of clusters. Conversely, too few regions may prevent the upper layer from identifying demand-supply imbalances, causing the rebalancing to rely heavily on coverage control. Therefore, an interesting direction for future research is to investigate how to define the regions, as this can significantly influence the performance of the proposed framework.

\begin{appendices}
\section{List of Hyperparameters}\label{append_hyper}
\cref{tab:hyper} lists the values of hyperparameters of our proposed algorithm, for both upper- and lower-layer controllers. Unless stated in the text, the specified value in this list is applied across all results.

\begin{table*}[htb]
\centering
\begin{tabular}{|c|c|c|}
\hline
\textbf{Hyperparameter} & \textbf{Value} & \textbf{Description} \\ 
\hline
$R$     &5    & number of clustered regions\\
\hline
$T_d$  & 3000 & length of historical data for reconstructing Hankel matrices \\ 
\hline
$T_{ini}$ & 35 & length of initial inputs and outputs \\ 
\hline
$\lambda_g$ & 1000 & regularization weight of $g$ \\ 
\hline
$\lambda_y$ & 0.01 & regularization weight of $y$ \\ 
\hline
$\Phi^O$ & $[0.06, 0.35, 0.22, 0.29, 0.08]$ & Regional trip origin probability vector \\ 
\hline
$\Phi^D$ & $[0.16, 0.28, 0.17, 0.27, 0.12]$ & Regional trip destination probability vector \\ 
\hline
$\mathcal{Q}$ & $col(diag(\Phi^O))$ & State weight matrix \\ 
\hline
$\mathcal{R}$ & $col(\{ L^{IJ}\})$  & Control weight matrix \\ 
\hline

$\{L^{IJ}\}$ & $
\begin{pmatrix}
 0     &  1.83 &  2.07 &  5.35 &  3.94\\
 1.58  &  0    &  2.08 &  4.98 &  4.94\\
 2.56  &  2.20 &   0   &  2.70 &  2.15\\
 6.08  &  4.77 &  2.19 &  0    &  2.73\\
 4.23  &  4.61 &  1.70 &  2.47 &  0
\end{pmatrix}$   & Average rebalancing trip length (unit:km) from Region $I$ to Region $J$ \\ 
\hline
$\alpha$ & $150$ & tunning parameter of $\mathcal{Q}$ \\ 
\hline
$N$ & $10$ & Upper-layer time horizon \\ 
\hline

$T_u$ & $10 min$ & Time step of the upper-layer controller\\
\hline
$T_l$ & $30 s$ & Time step of the lower-layer controller\\
\hline
$T_m$ & $1 min$ & Matching time threshold\\
\hline
$T_w$ & $4 min$ & Waiting time threshold\\
\hline

\end{tabular}
\caption{List of hyperparameter values and descriptions}
\label{tab:hyper}
\end{table*}

\end{appendices}
\section*{Dataset Availability}
The passenger request dataset is available via \textcolor{blue}{\url{https://doi.org/10.5281/zenodo.8287367}}

\section*{Acknowledgments}
This research is supported by NCCR Automation, a National Centre of Competence in Research funded by the Swiss National Science Foundation (grant number 51NF40\_180545).

\bibliographystyle{IEEEtran}
\bibliography{ref}

\begin{IEEEbiography}[{\includegraphics[width=1in,height=1.25in,clip,keepaspectratio]{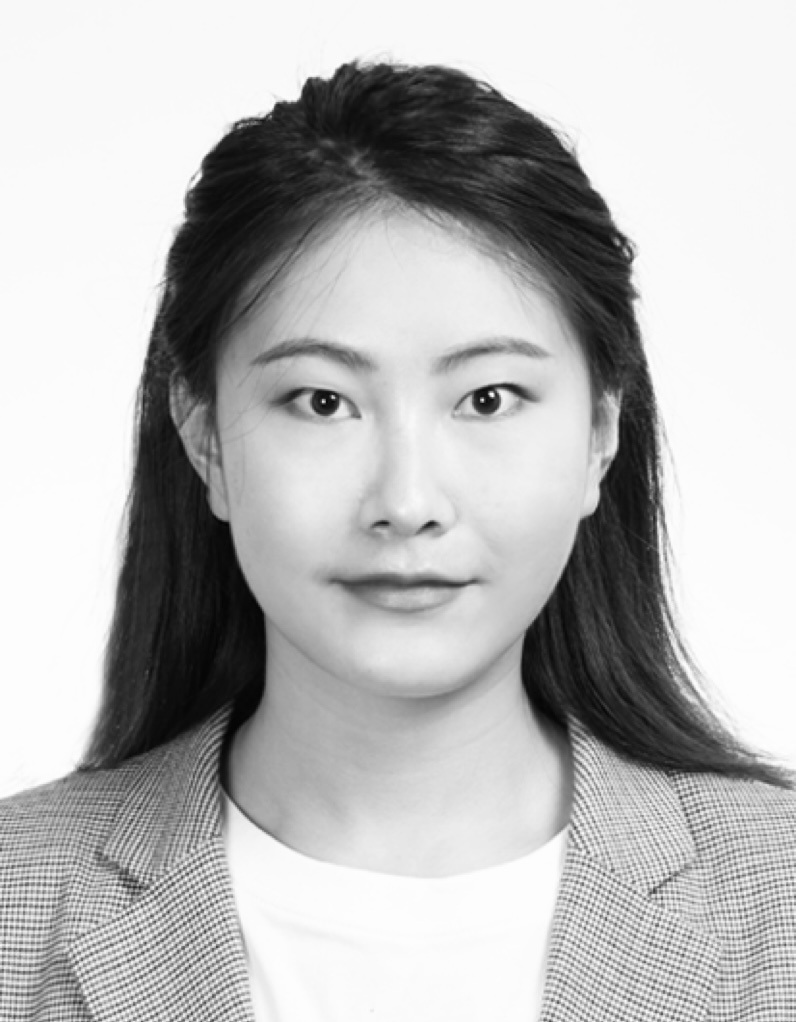}}]{Pengbo Zhu} received the B.Sc. degree in Automation in 2017 and M.Sc. degree in Control Science and Engineering in 2019 from Harbin Institute of Technology. Currently, she is a doctoral assistant at the Urban Transport Systems Laboratory (LUTS) of EPFL. Her research interests are the application of control algorithms in urban transportation systems, and the optimization of emerging mobility systems.

\end{IEEEbiography}

\begin{IEEEbiography}[{\includegraphics[width=1in,height=1.25in,clip,keepaspectratio]{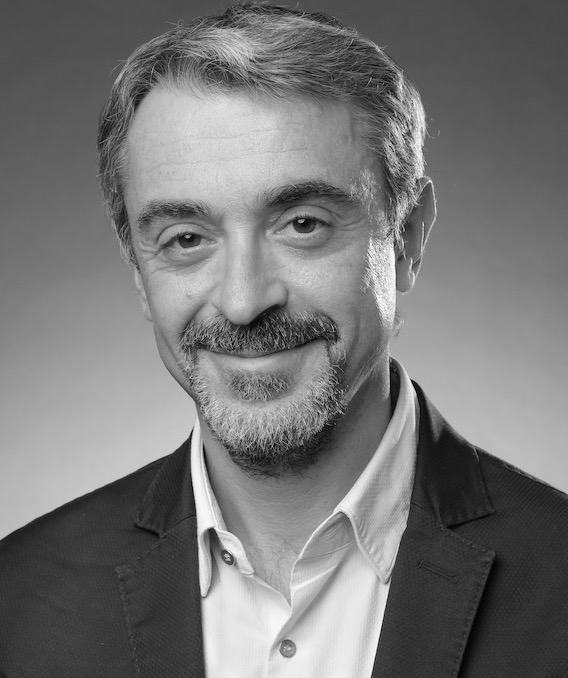}}]{Giancarlo Ferrari-Trecate}(SM’12) received the Ph.D. degree in Electronic and Computer Engineering from the Universita' Degli  Studi di Pavia in 1999. Since September 2016, he has been Professor at EPFL, Lausanne, Switzerland. In the spring of 1998, he was a Visiting Researcher at the Neural Computing Research Group, University of Birmingham, UK. In the fall of 1998, he joined the Automatic Control Laboratory, ETH, Zurich, Switzerland, as a Postdoctoral Fellow. He was appointed Oberassistent at ETH in 2000. In 2002, he joined INRIA, Rocquencourt, France, as a Research Fellow. From March to October 2005, he was a researcher at the Politecnico di Milano, Italy. From 2005 to August 2016, he was Associate Professor at the Dipartimento di Ingegneria Industriale e dell'Informazione of the Universita' degli Studi di Pavia. His research interests include scalable control, microgrids, machine learning, networked control systems, and hybrid systems. Giancarlo Ferrari Trecate is the founder of the Swiss chapter of the IEEE Control Systems Society and is currently a member of the IFAC Technical Committees on Control Design and Optimal Control. He has been serving on the editorial board of Automatica for nine years and of Nonlinear Analysis: Hybrid Systems.
\end{IEEEbiography}

\begin{IEEEbiography}[{\includegraphics[width=1in,height=1.25in,clip,keepaspectratio]{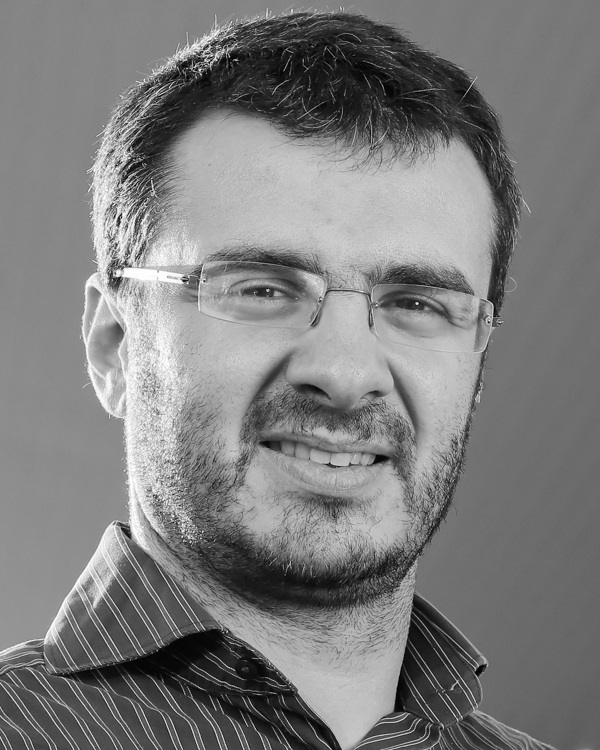}}]{Nikolas Geroliminis} is a full professor at EPFL and the head of the Urban Transport Systems Laboratory (LUTS). He has a diploma in Civil Engineering from the National Technical University of Athens (NTUA) and an M.Sc. and Ph.D. in civil engineering from University of California, Berkeley. He is a member of the Transportation Research Board's Traffic Flow Theory Committee.  He also serves as an Associate Editor in Transportation Research, part C, Transportation Science and IEEE Transactions on ITS and in the editorial board of Transportation Research, part B, Journal of ITS and of many international conferences. His research interests focus primarily on urban transportation systems, traffic flow theory and control, public transportation and logistics, on-demand transportation, optimization and large scale networks. He is a recent recipient of the ERC starting grant ``METAFERW: Modeling and controlling traffic congestion and propagation in large-scale urban multimodal networks''.
\end{IEEEbiography}

\end{document}